\documentclass[a4paper,11pt]{article}
\usepackage{pos}
\usepackage{setspace}
\usepackage{amsmath}

\title{The muon deficit problem: a new method to calculate the muon rescaling factors and the Heitler-Matthews $\beta$ exponent}
\ShortTitle{The muon deficit problem: a new method...}

\author[a]{Kevin Almeida Cheminant}
\author[a]{Dariusz Góra}
\author[a]{Nataliia Borodai}
\author[b]{Ralph Engel}
\author[b]{Tanguy Pierog}
\author*[a]{Jan Pękala}
\author[b]{Markus Roth}
\author[a]{Jarosław Stasielak}
\author[b]{Michael Unger}
\author[b]{Darko Veberič}
\author[a]{Henryk Wilczyński}

\affiliation[a]{Institute of Nuclear Physics PAS,\\
  Radzikowskiego 152, Kraków, Poland}

\affiliation[b]{Karlsruhe Institute of Technology, Institute for Astroparticle Physics,\\
Karlsruhe, Germany}

\emailAdd{jan.pekala@ifj.edu.pl}

\abstract{Simulations of extensive air showers using current hadronic interaction models predict too small numbers of muons compared to events observed in the air-shower experiments, which is known as the muon-deficit problem. In this work, we present a new method to calculate the factor by which the muon signal obtained via Monte-Carlo simulations must be rescaled to match the data, as well as the $\beta$ exponent from the Heitler-Matthews model which governs the number of muons found in an extensive air shower as a function of the mass and the energy of the primary cosmic ray. This method uses the so-called $z$ variable (difference between the total reconstructed and the simulated signals), which is connected to the muon signal and is roughly independent of the zenith angle, but depends on the mass of the primary cosmic ray. Using a mock dataset built from QGSJetII-04, we show that such a method allows us to reproduce the average muon signal from this dataset using Monte-Carlo events generated with the EPOS-LHC hadronic model, with accuracy better than 6\%. As a consequence of the good recovery of the muon signal for each primary included in the analysis, also the $\beta$ exponent can be obtained with accuracy of less than 1\% for the studied system. Detailed simulations show a dependence of the $\beta$ exponent on hadronic interaction properties, thus the determination of this parameter is important for understanding the muon deficit problem.}

\FullConference{%
  *** 27th European Cosmic Ray Symposium - ECRS ***\\
  *** 25-29 July 2022 ***\\
  *** Nijmegen, the Netherlands ***
}

\begin{document}
\maketitle

\section{Introduction}

In this work we examine the deficit of muons in simulations of extensive air showers -- the muon number observed in data at energies around $10^{19}$ eV is 30\% to 60\% higher than in corresponding simulations \cite{3}. The analyses of the observational data depend on simulations, therefore this deficit will systematically influence the results, e.g. studies of the composition of the primary cosmic rays based on observed muon numbers produce results shifted towards heavier elements as compared with analyses based on measurements of $X_\text{max}$ (i.e. the atmospheric depth at which the air shower reaches its maximum size) \cite{4}.

To study the muon deficit we use a top-down (TD) method -- its chain of simulations and reconstructions enable us to calculate signals in the fluorescence (FD) and surface detectors (SD) of the Pierre Auger Observatory \cite{5} (Auger). For each observed shower, starting with a large number of simulated air showers with varying initial conditions, we select the one which has a longitudinal profile most similar to the profile of the observed shower (the reference profile). As a result of the simulation-reconstruction chain we get an event, with complete information about the distributions of the signals in the detectors (including information on the specific components that contribute to these signals) -- these signals can then be compared with their reference counterparts. Since the results of the simulations depend on the properties of the hadronic interaction models that are included in the simulation software, by comparing the simulations with corresponding observational results we should be able to verify these models at energies exceeding those available in any man-made accelerators. We expect that we will gain new information, which will enable improvment of the interaction models, and in this way we are also able to reduce the discrepancy between the observations and simulations.

\begin{figure}[!b]
\begin{center}
\includegraphics[scale=0.9]{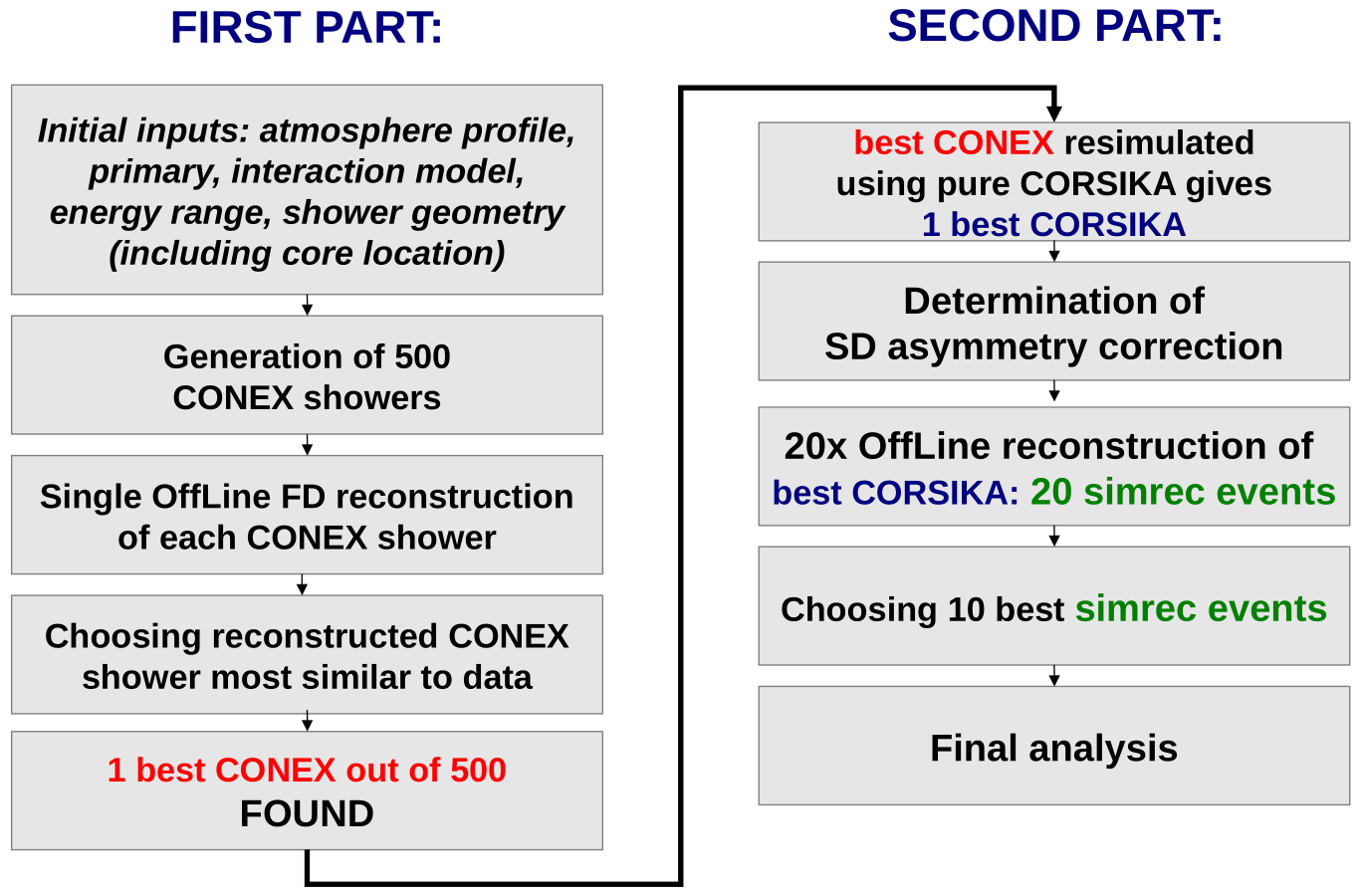}
\caption{A scheme of top-down simulation-reconstruction chain.}
\label{f0}
\end{center}
\end{figure}

\section{Top-Down method}

\begin{figure}[!t]
\includegraphics[scale=0.595]{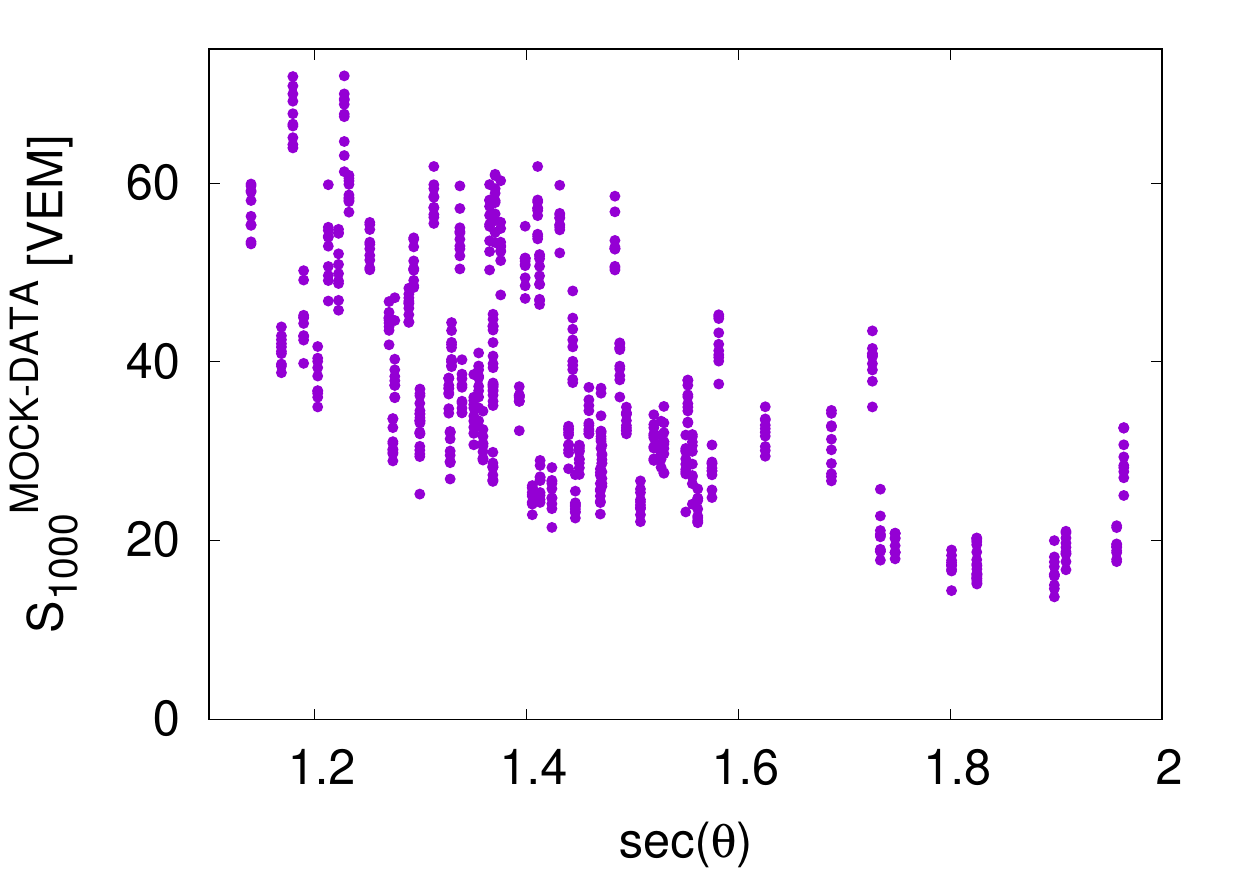}\includegraphics[scale=0.595]{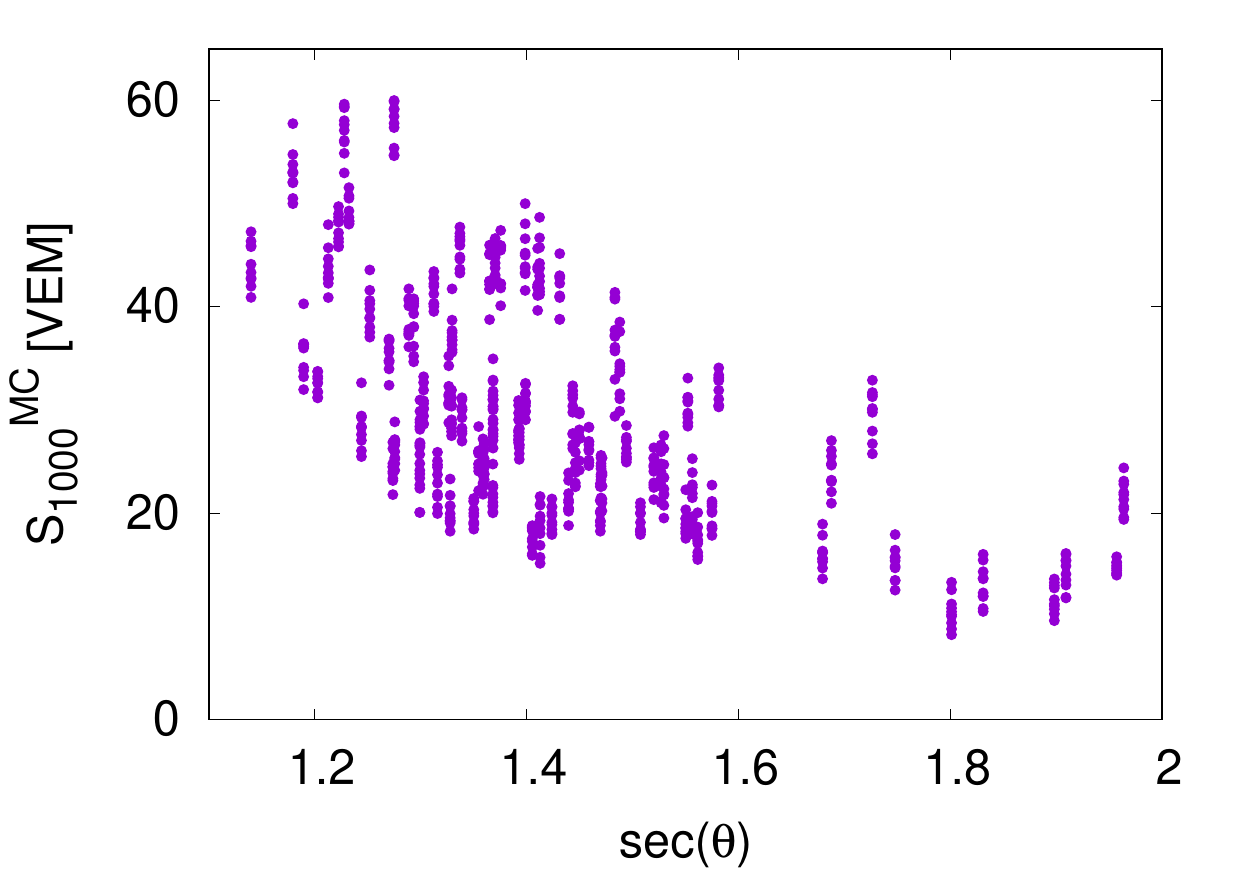}
\caption{\emph{Left}: the total MOCK-DATA signal $S_{1000}$ as a function of zenith angle $\theta$, at a reference distance of 1000\,m from the shower core; air showers of $10^{19}$ eV energy with iron primaries and simulated with \mbox{EPOS-LHC} model. \emph{Right}: the total MC signal, for proton air showers of $10^{19}$ eV simulated with \mbox{QGSJetII-04}.}
\label{f1}
\end{figure}

In this work we use a set of simulated air showers (MOCK-DATA) as a reference -- by starting with showers with exactly known properties, and reconstructing them independently we are able to demonstrate the capabilities of the TD method. The TD method we use is an improved version of the simulation-reconstruction chain described in Ref. \cite{3}. We have replaced the SENECA software \cite{7}, that was used in simulations, with CORSIKA \cite{6}, which was extensively developed and is currently more reliable and accurate. The details of our TD method are described in Refs.~\cite{8,8a}. Fig.~\ref{f0} presents the scheme of the TD chain, that for each of the reference shower generates a set of simulated showers, from which the best one is selected (based on $\chi^2$ calculated between reference and simulated profile). As a final result of this method we get 10 reconstructions of this shower (performed with the Offline software \cite{9}) which are used in the following analysis. The TD calculations were done using two of the most up-to-date nuclear interaction models: we use 78 events for EPOS-LHC \cite{1} and 89 for QGSJetII-04 \cite{2}. Simulations with EPOS-LHC and iron primaries were chosen as the MOCK-DATA showers (Fig.~\ref{f1}) as they produce muon signals most similar to the observed ones at the energy of $10^{19}$ eV \cite{10}. The QGSJetII-04 simulations that are used to reconstruct the muon signals of the MOCK-DATA set were repeated for four different types of primary cosmic rays: proton, helium, nitrogen and iron.

\begin{figure}[!t]
\includegraphics[scale=0.595]{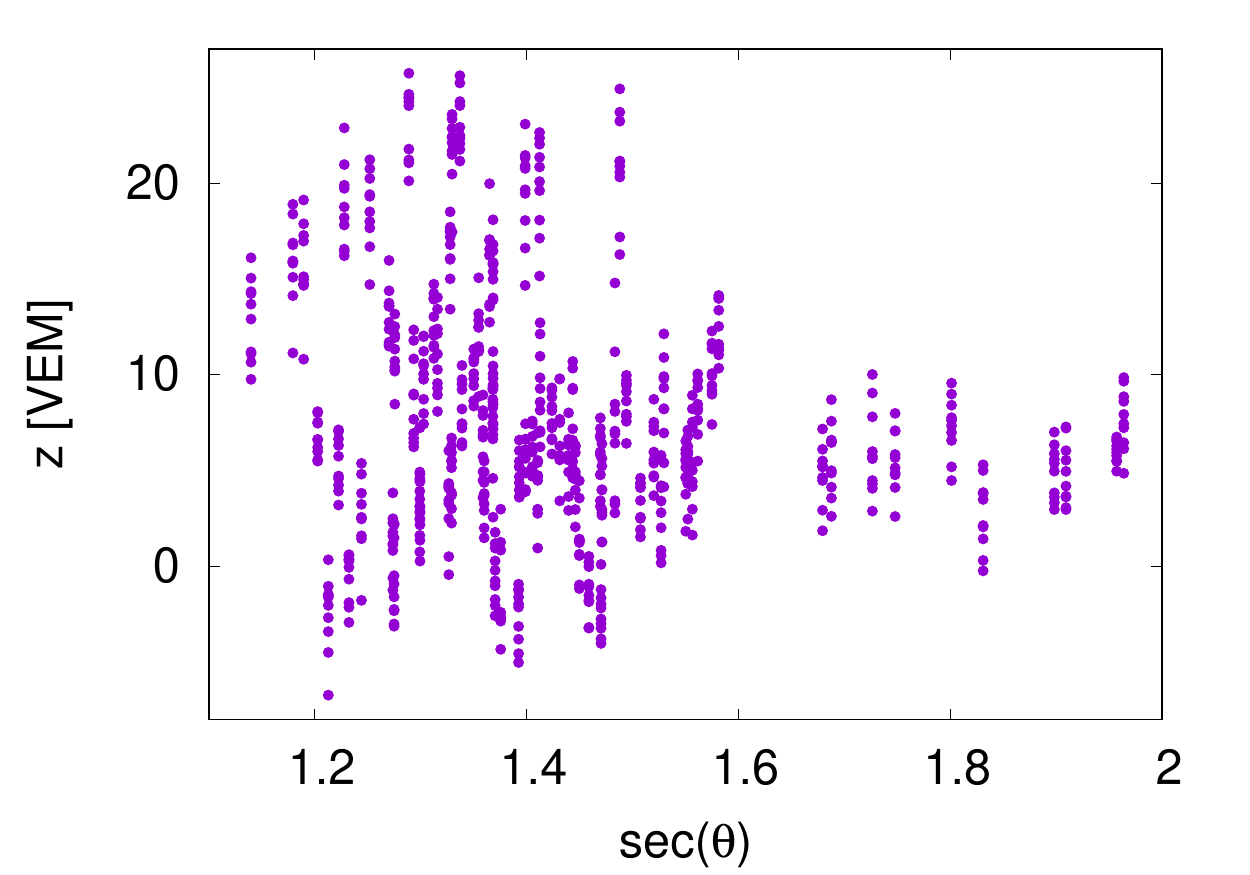}\includegraphics[scale=0.595]{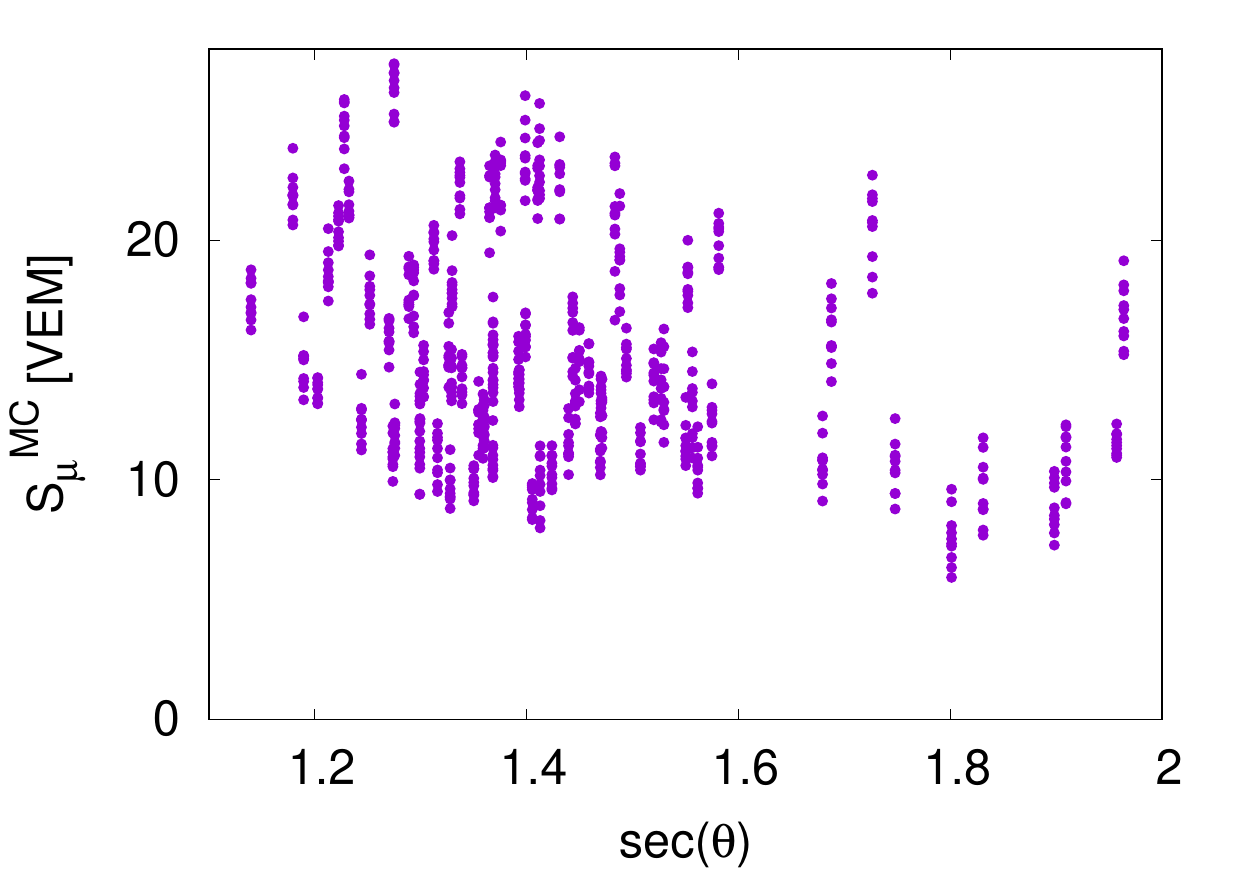}
\caption{\emph{Left}: the variable $z$. \emph{Right}: $S_{\upmu}$ for proton air showers of $10^{19}$ eV, simulated with QGSJetII-04.}
\label{f2}
\end{figure}

Almost the whole of the total signal at the ground comes from the electromagnetic and muonic components \cite{3} -- we will denote these contributions $S_\text{EM}$ and $S_{\upmu}$, respectively. Other components of the shower \cite{11} do not contribute significantly. Any discrepancy between corresponding observed and simulated values can be described by linear scaling factors: $R_E$ for the electromagnetic part of an air shower, and $R_{\upmu}$ for the muonic part. With these assumptions the ground signal at 1000\,m from the axis ($S_{1000}$) for MOCK-DATA and simulated air shower $j$ can be respectively written as:

\begin{equation}
S^{\text{MC}}_{1000,j} \equiv S^{\text{MC}}_{\text{EM},j} +  S^{\text{MC}}_{\upmu,j},
\label{e1}
\end{equation}
\begin{align}
S^{\text{MOCK-DATA}}_{1000,j}(R_{E},R_\upmu)_{j} &\equiv S^{\text{MOCK-DATA}}_{\text{EM},j} +  S^{\text{MOCK-DATA}}_{\upmu,j}, \nonumber \\
S^{\text{MOCK-DATA}}_{\text{EM},j} &= R_{E} \, S^{\text{MC}}_{\text{EM},j}, \label{e2} \\
S^{\text{MOCK-DATA}}_{\upmu,j} &= R_\upmu \, S^{\text{MC}}_{\upmu,j}.
\nonumber
\end{align}

The comparison of simulations and measurements performed in Ref.~\cite{3} has shown that the value of $R_E$ is very close to one. In our TD method, when we select the simulated shower through comparison with the longitudinal profile of the MOCK-DATA shower, we increase our confidence in the accuracy of the simulation even more. Therefore in our analysis we assume $R_E$ = 1.

Rather than using the ground signal directly, we use the difference between MOCK-DATA and the MC signal as the main observable:
\begin{center}
$z_{j} \equiv S_{1000,j}^{\text{MOCK-DATA}} - S_{1000,j}^{\text{MC}}$.
\end{center}
This difference explicitly shows the discrepancy between observations and simulations. The variable $z$ is roughly independent of the zenith angle -- as is shown for example in Fig. \ref{f2} (left), the distribution of $z$ only slightly depends on the zenith angle, as opposed to the $S_{1000}$ signal shown in Fig. \ref{f1}. The largest advantage of using the variable $z$ becomes apparent after comparing Eqs. \eqref{e1} and \eqref{e2}. By simple subtraction, and assuming $R_E$ = 1, we get
\begin{equation}
S^{\text{MC}}_{\upmu,i,j} = \frac{z_{i,j}}{R_{\upmu,i,j}-1}.
\label{e3}
\end{equation}
We see that the simulated muon signal is proportional to the difference between observed and MC signal $z$, and the ratio depends on the muon scaling factor. The average signal difference, $\langle z_i \rangle$,  depends on the primary type, and we have to account for this effect (additional index $i \in$ \{p, He, N, Fe\}). Air showers initiated by heavier primary cosmic rays of the same energy contain larger number of muons, and correspondingly larger muon signals at the ground are expected \cite{13}, therefore muon scaling factors $R_{\upmu,i}$ should be studied separately at different primary masses.

The average muon signal as a fraction $g_{\upmu,i}(\theta)$ of the total signal at the ground has been calculated in previous analyses. This fraction depends on the shower zenith angle and the type of the primary cosmic ray, and only slightly on different hadronic interaction models \cite{10}. For our analysis the $g_{\upmu,i}(\theta)$ was obtained from the analysis of simulations of muon signal in stations located at a distance of 1000\,m from the shower core for showers with energies of $10^{18.5}$ to $10^{19}$ eV.

We know the values of $z_{i,j}$ and the $S_{1000}$, and using the parameterization $g_{\upmu,i}(\theta)$, we can get the muon scaling factor even for an individual air shower $j$,
\begin{equation}
R_{\upmu,i,j}=1+\frac{z_{i,j}}{g_{\upmu,i}(\theta) \, S^{\text{MC}}_{1000,i,j}},
\label{e4}
\end{equation}
where $S^{\text{MC}}_{\upmu,i,j} \equiv g_{\upmu,i}(\theta) \, S^{\text{MC}}_{\text{1000},i,j}$.

\section{Muon rescaling factors}

The average reconstructed muon signals $\langle S^{\text{MC}}_{\upmu,i} \rangle$ are listed in Table \ref{t1}. For all primaries the QGSJetII-04 model predicts smaller muon signals than simulations with the EPOS-LHC model -- that was expected based on the results from Ref.~\cite{16}. The difference between both models is in the number of muons, but also in the energy spectrum of the muons (e.g.\ EPOS-LHC predicts more muons with lower energies) -- this leads to a difference of the muon signal at the ground between simulations done with both models. From Table \ref{t1}, we can calculate the average ratio of the muon signals as $r = S^{\text{MC-EPOS}}_{1000,\upmu,i} / S^{\text{MC-QGSJet.}}_{1000,\upmu,i} \simeq$ 1.11 $\pm$ 0.04. The average value of individual factors $R_{\upmu,i}$ is almost independent of the zenith angle and decreases with growing primary mass. 

Table \ref{t2} presents values of the total muon signal $S^{\text{MOCK-DATA}}_{\upmu} \equiv \langle R_{\upmu,i} \rangle \, \langle S^{\text{MC}}_{1000,\upmu,i} \rangle$. $S^{\text{MC-True}}_{\upmu}$ is known from the initial MOCK-DATA simulations, and $S^{\text{MOCK-DATA}}_{\upmu}$ is the corresponding estimate obtained in the TD procedure. The reconstructed signal is by only 2 to 4\% different from the muon signal from the MOCK-DATA set ($S^{\text{MC-True}}_{\upmu}$= 23.09 VEM) -- this demonstrates the accuracy of the presented method.

\begin{table}
\begin{center}
\begin{tabular}{ |c c c c c c c| } 
 \hline
 primary type $i$ & model & $\langle S^{\text{MC}}_{\upmu,i} \rangle$ & st. dev. & model & $\langle S^{\text{MC}}_{\upmu,i} \rangle$ & st. dev. \\
  &  & [VEM] & [VEM] &  & [VEM] & [VEM] \\
 \hline
 p & EPOS-LHC & 16.89 $\pm$ 0.31 & 5.1 & QGSJetII-04 & 15.05 $\pm$ 0.3 & 4.3 \\
 He &  & 18.74 $\pm$ 0.42 & 5.7 &  & 16.82 $\pm$ 0.4 & 4.7 \\
 N &  & 20.67 $\pm$ 0.37 & 5.9 &  & 18.96 $\pm$ 0.4 & 5.1 \\
 Fe &  & 23.09 $\pm$ 0.42 & 6.4 &  & 21.08 $\pm$ 0.4 & 5.7 \\
 \hline
\end{tabular}
\end{center}
\caption{Results obtained using the TD method: the mean muon signal $S_{1000,\upmu,i}$ and its standard deviation for EPOS-LHC and QGSJetII-04.}
\label{t1}
\end{table}

\begin{table}
\begin{center}
\begin{tabular}{ |c c c c c c| } 
 \hline
 primary type $i$ & $\langle R_{\upmu,i} \rangle$ & $\sigma_{\langle R_{\upmu,i} \rangle}$ & $\langle S^{\text{MC}}_{\upmu,i} \rangle$ & $\langle R_{\upmu,i} \rangle \, \langle S^{\text{MC}}_{\upmu,i} \rangle$ & $k$ \\
  &  &  & [VEM] & [VEM] & [\%] \\
 \hline
 p & 1.57 $\pm$ 0.01 & 0.27 & 15.05 & 23.63 & 2.3 \\
 He & 1.42 $\pm$ 0.01 & 0.26 & 16.82 & 23.88 & 3.4 \\
 N & 1.26 $\pm$ 0.01 & 0.21 & 18.96 & 23.89 & 3.5 \\
 Fe & 1.14 $\pm$ 0.01 & 0.18 & 21.08 & 24.00 & 3.9 \\
 \hline
\end{tabular}
\end{center}
\caption{The mean values of the muon rescaling parameters $R_{\upmu,i}$ calculated from Eq. \eqref{e4}, and their standard deviations $\sigma_{\langle R_{\upmu,i} \rangle}$, mean values of the muon ground signal $S^{\text{MC}}_{1000,\upmu,i}$ from simulations with QGSJetII-04 model, reconstructed muon ground signal predicted for the MOCK-DATA set, and the ratio $k \equiv (\langle R_{\upmu,i} \rangle \, \langle S^{\text{MC}}_{1000,\upmu} \rangle - S^{\text{MC-True}}_{\upmu}) / S^{\text{MC-True}}_{\upmu}$.}
\label{t2}
\end{table}

\section{Calculating the \texorpdfstring{$\beta$}{beta} exponent}

Observations of air showers, and also simulations, demonstrated that the number of muons $N_{\upmu}$ grows almost linearly with the shower energy $E$, and it also increases with a small power of the primary mass $A$. These relations can be reproduced in the framework of the Heitler-Matthews model of hadronic air showers \cite{13}. This model predicts
\begin{center}
$N_{\upmu} = A(\frac{E/A}{\epsilon_{\text{c}}^{\pi}})^{\beta}$,
\end{center}
where $\beta \approx 0.9$. For any fixed energy it describes how the muon number depends on the primary mass $N_{\upmu}=N_{\upmu,\text{p}} \, A^{1-\beta}$ (the $N_{\upmu,\text{p}}$ is the number of muons of a proton shower; $\epsilon_{\text{c}}^{\pi}$ is the critical energy at which pions decay into muons).

Simulations have shown that $\beta$ depends on various properties of hadronic interaction (e.g. multiplicity, charge ratio, baryon anti-baryon pair production \cite{14}). Therefore, estimating the $\beta$ exponent would be helpful in constraining the parameters of hadronic interactions and improving the accuracy of models.

We can assume that the average muon signal $S_{\upmu}$ is proportional to muon number $N_{\upmu}$. With this we can calculate the logarithm of the muon number $N_{\upmu,i}$ for primary $i$ and iron ($A=56$; we choose it as a reference), and we get: $\beta_{i}=1-\frac{\ln \langle S^{\text{MC}}_{\upmu,\text{Fe}}\rangle -\ln \langle S^{\text{MC}}_{\upmu,i} \rangle }{\ln 56-\ln A_{i}}$. The $\beta$ exponent calculated using the muon signals listed in the Table \ref{t1}  is $\sim$0.92 -- this is close to the values reported in Ref~\cite{15}: $\beta$ = 0.927 for EPOS-LHC and $\beta$ = 0.925 for QGSJetII-04, which is another validation of our method. We can also use the reconstructed muon signals for any primary $i$, $S^{\text{MOCK-DATA}}_{\upmu,i} \equiv  \langle R_{\upmu,i} \rangle \, \langle S^{\text{MC}}_{1000,\upmu,i} \rangle$, to calculate the $\beta$ exponent. In this case the $\beta_i$ exponent is given by 
\begin{center}
$\beta_{i}=1-\frac{\ln \langle S^{\text{MOCK-DATA}}_{\upmu,\text{Fe}}\rangle -\ln \langle S^{\text{MOCK-DATA}}_{\upmu,i} \rangle }{\ln 56-\ln A_{i}}=1-\frac{\ln\langle(R_{\upmu,\text{Fe}}\rangle \langle S^{\text{MC}}_{\upmu,\text{Fe}}\rangle) -\ln (\langle R_{\upmu,i}\rangle \langle S^{\text{MC}}_{\upmu,i}\rangle) }{\ln 56-\ln A_{i}}$.
\end{center}

\begin{figure}[!t]
\includegraphics[scale=0.595]{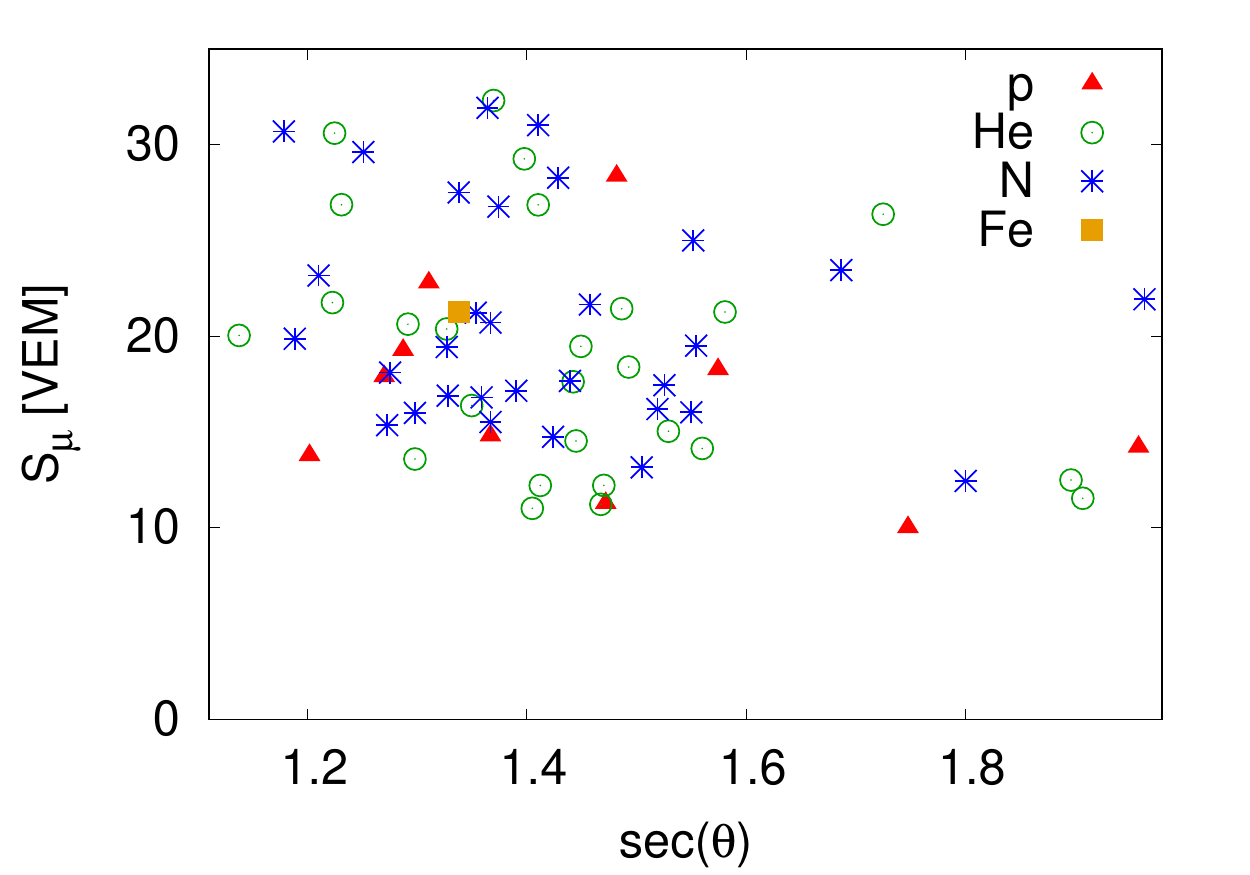}\includegraphics[scale=0.595]{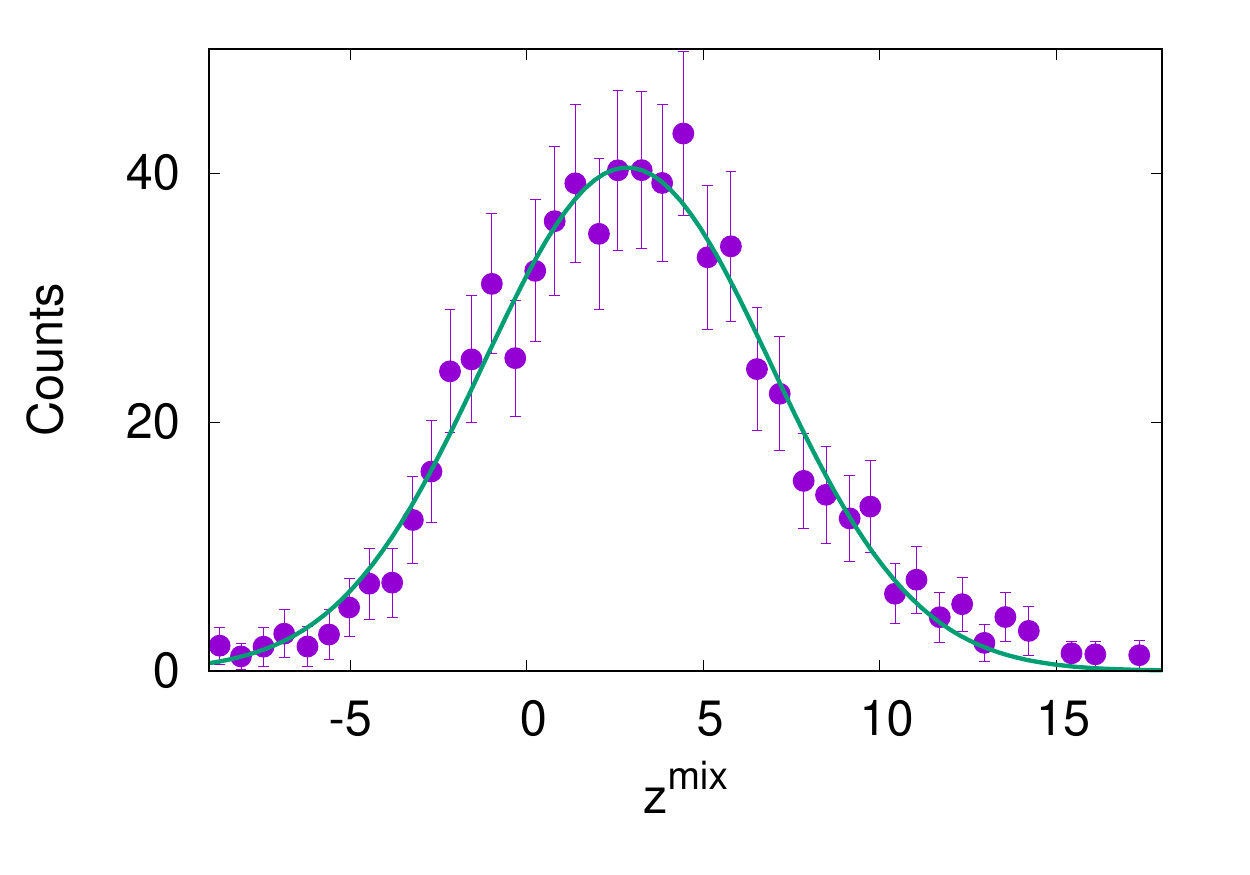}
\caption{\emph{Left}: the $S_{\upmu,i}$ as a function of zenith angle. The distribution of events (68 in total) follows the composition reported by Auger for EPOS-LHC \cite{17}. \emph{Right}: the $z^{\text{mix}}$ distribution for MOCK-DATA set. The green line shows an example of a Gaussian fit given by Eq. \eqref{e5}.}
\label{f4a}
\end{figure}

To calculate the $\beta_i$ exponent for a set of air showers with a predefined composition we perform the following operations. First we generate a new MOCK-DATA set, shown in Fig.~\ref{f4a} (left). This MOCK-DATA set is constructed so that it contains a sample of events with fractions of primary types ($f_i$) that reflect the composition observed by Auger at $10^{19}$ eV ($f_{\text{p}} \simeq$ 15\%, $f_{\text{He}} \simeq$ 38\%, $f_{\text{N}} \simeq$ 46\%, $f_{\text{Fe}} \simeq$ 1\% \cite{17}). Also the distribution of the zenith angles of the events was selected so that it follows the distribution of the TD simulations. The average muon signal for all primaries is also similar (within $\pm$0.5 VEM) to our previous results with the EPOS-LHC model. Using this new MOCK-DATA set we calculate for each of the events the $z^{\text{mix}}$ variable, $z^{\text{mix}}_{j} \equiv S^{\text{MOCK-DATA}}_{1000,j}-\sum_{\text{i} \in \{\text{p,He,N,Fe}\}}f_{i}\,S^{\text{MC}}_{1000,i,j}.$ The distribution of the $z^{\text{mix}}$ is shown in Fig.~\ref{f4a} (right). To this $z^{\text{mix}}$ histogram we fit the Gaussian function given by
\begin{equation}
P(A,\sigma, {\bf R}^{\text{fit}}_{\upmu})=B\exp(-(z^{\text{mix}}-\langle z^{\text{mix}} \rangle )^2/2\sigma^2),
\label{e5}
\end{equation}
\begin{equation}
\langle z^{\text{mix}} \rangle=\sum_{\text{i}\in \{\text{p,He,N,Fe\}}} f_{i}\, \langle S^{\text{MC}}_{\upmu,i}\rangle (R^{\text{fit}}_{\upmu,i}-1).
\label{e6}
\end{equation}
The fitting parameters are: the amplitude $B$, the standard deviation $\sigma$ and the rescaling parameters ${{\bf R}^{\text{fit}}_{\upmu}} = \{R^{\text{fit}}_{\upmu,\text{p}}, R^{\text{fit}}_{\upmu,\text{He}}, R^{\text{fit}}_{\upmu,\text{N}}, R^{\text{fit}}_{\upmu,\text{Fe}} \}$. According to Eq. \eqref{e3} the average of the muon signal is proportional to the $z^{\text{mix}}$, and $R^{\text{fit}}_{\upmu,i} \, \langle S^{\text{MC}}_{\upmu,i} \rangle$ are the contributions from the corresponding primaries $i$ to the total muon signal.

To find the most likely solutions we have fitted the Gaussian function to $z^{\text{mix}}$ histogram using the TMinuit routine from the ROOT package. We have done many minimizations with different initial values of the muon scaling parameters -- we performed a scan for $1 \leq R_{\upmu,i} \leq 2$ with a step of 0.005. Some cuts were applied to select the correct solutions. One condition was $R_{\upmu,\text{p}} \langle S^{\text{MC}}_{\upmu,\text{p}} \rangle < R_{\upmu,\text{He}} \langle S^{\text{MC}}_{\upmu,\text{He}} \rangle < R_{\upmu,\text{N}} \langle S^{\text{MC}}_{\upmu,\text{N}} \rangle < R_{\upmu,\text{Fe}} \langle S^{\text{MC}}_{\upmu,\text{Fe}} \rangle$ -- from the physics of air showers we know that muon numbers should increase with the mass of the primary. Based on the Heitler-Matthews model it is also expected, that logarithm of the muon signal should increase linearly with logarithm of the primary mass, therefore corresponding linearity conditions were introduced:

\begin{center}
$\Big{|} \frac{\ln(R_{\upmu,\text{p}} \langle S_{\upmu,\text{p}}^{\text {MC}}\rangle)  -\ln(R_{\upmu,\text{He}} \langle S_{\upmu,\text{He}}^{\text{MC}}\rangle)}{\ln(A_{\text{p}})-\ln(A_{\text{He}})}   - \frac{\ln(R_{\upmu,\text{He}} \langle S_{\upmu,\text{He}}^{\text {MC}}\rangle)  -\ln(R_{\upmu,\text{N}} \langle S_{\upmu,\text{N}}^{\text{MC}}\rangle)}{\ln(A_{\text{He}})-\ln(A_{\text{N}})} \Big{|} <\epsilon$,

$\Big{|}\frac{\ln(R_{\upmu,\text{He}} \langle S_{\upmu,\text{He}}^{\text MC}\rangle)  - \ln(R_{\upmu,\text{N}} \langle S_{\upmu,\text{N}}^{\text{MC}}\rangle)}{\ln(A_{\text{He}})-\ln(A_{N})}   - \frac{\ln(R_{\upmu,\text{N}} \langle S_{\upmu,\text{N}}^{\text{MC}}\rangle)  - \ln(R_{\upmu,\text{Fe}} \langle S_{\upmu,\text{Fe}}^{\text{MC}}\rangle)}{\ln(A_{\text{N}})-\ln(A_{\text{Fe}})} \Big{|}<\epsilon$.
\end{center}

\begin{figure}[!t]
\includegraphics[scale=0.595]{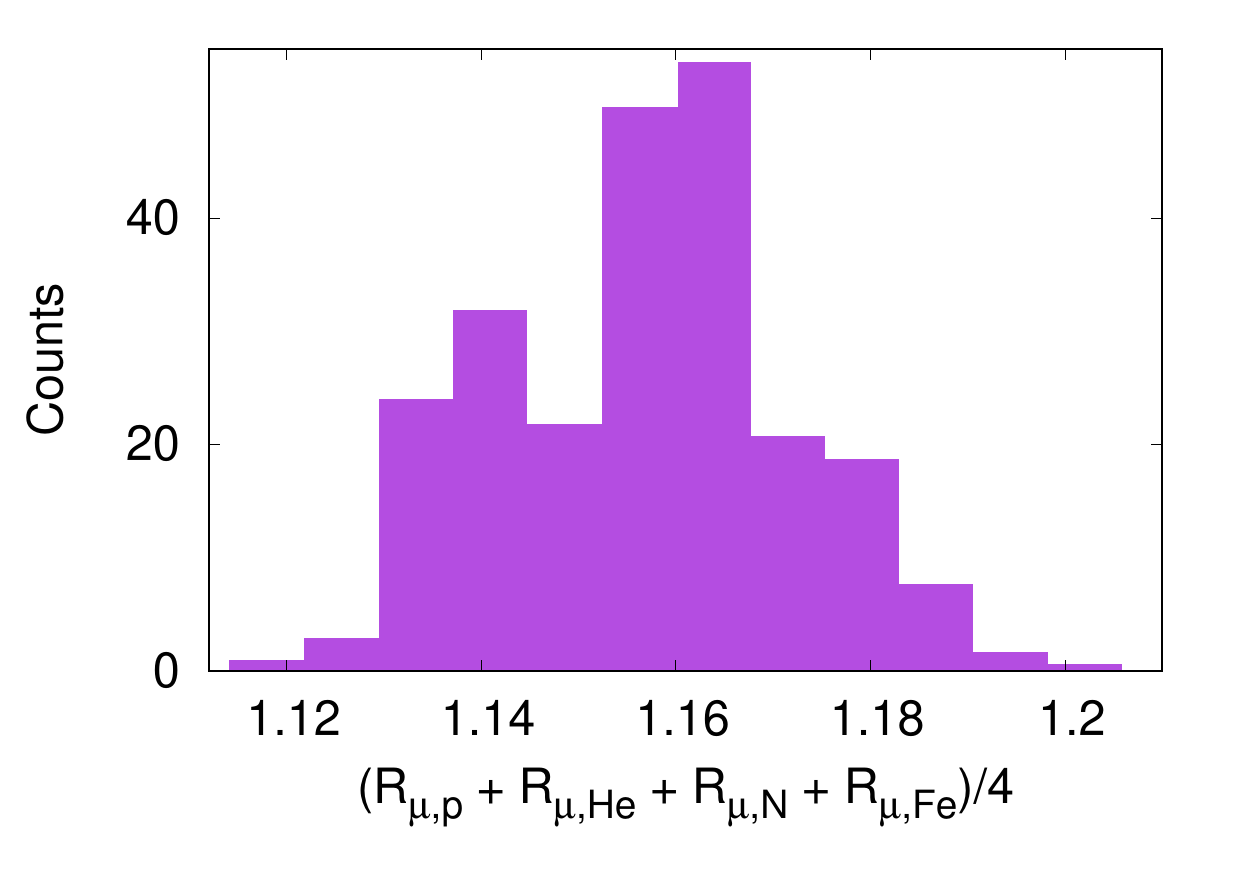}\includegraphics[scale=0.595]{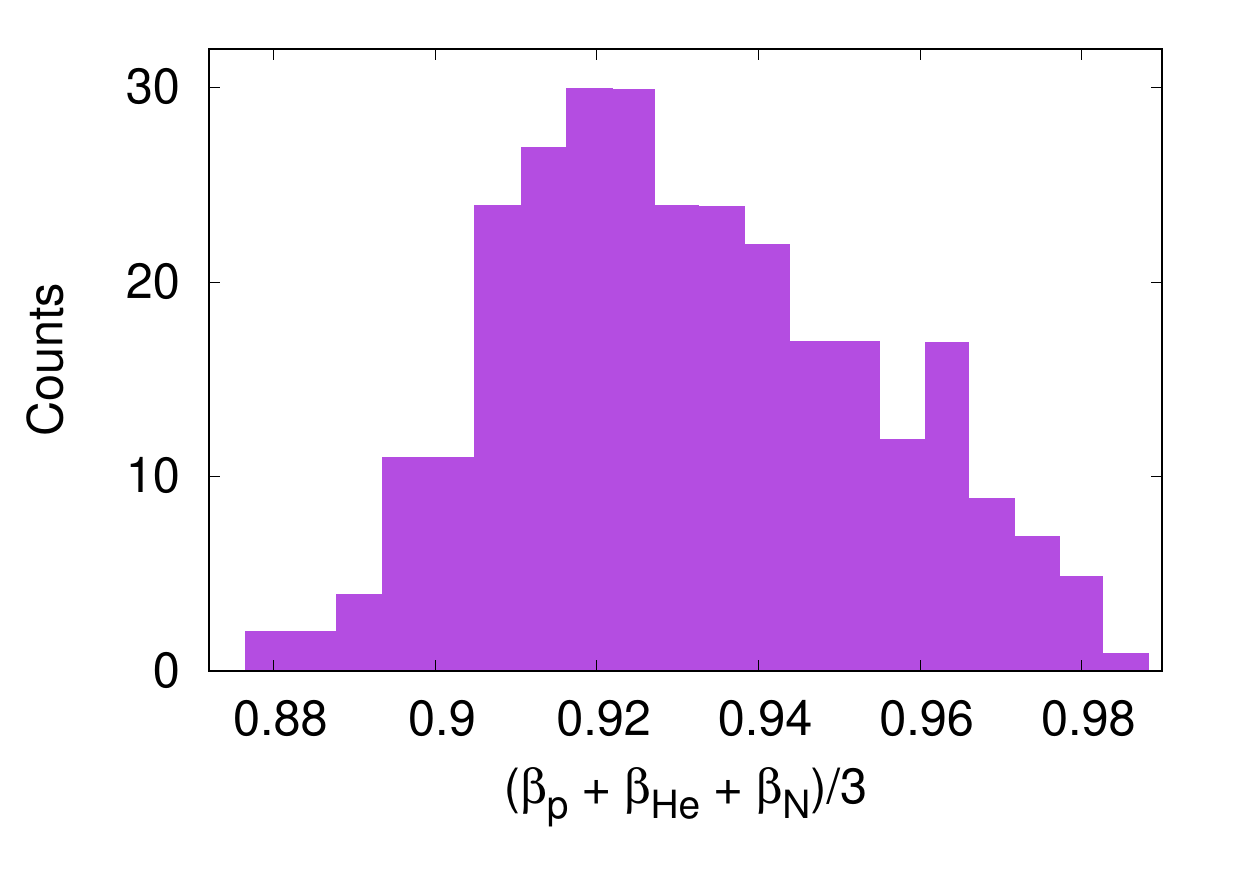}
\caption{The average muon rescaling factor (left) and the average $\beta_i$ exponent distribution (right) for events passing the selection cuts are shown.}
\label{f4b}
\end{figure}

The results are shown in Fig.~\ref{f4b}. We see that this method can reproduce the ratio of the muon signals of simulations using EPOS-LHC and QGSJetII-04 within $\sim$5\%: the ratio for MC-true is $r$~=~1.11 $\pm$ 0.04 and the average reconstructed ratio (left part of Fig.~\ref{f4b}) is 1.16 $\pm$ 0.02. The difference is caused by the fact that the signal for the MOCK-DATA set is not exactly equal to the one for EPOS-LHC (Table \ref{t1}). We also recover the $\beta$ parameter (average $\simeq$ 0.92) for the studied set.

\section{Summary}

This work presents a new method that can be used to determine the muon scaling factors, and also to estimate the $\beta$ exponent. This method has been tested with results from TD reconstructions of a set of air showers with the energy about $10^{19}$ eV. The method allows us to calculate the average muon signal for any set of air showers. Knowing this component is crucial for all analyses of air shower observations, and indirectly can also be used to improve hadronic interaction models.

\acknowledgments
The authors are very grateful to the Pierre Auger Collaboration for providing the tools necessary for simulation for this contribution. We want to acknowledge support in Poland from National Science Centre grant No.~2016/23/B/ST9/01635, grant No.~2020/39/B/ST9/01398 and from the Ministry of Science and Higher Education grant No.~DIR/WK/2018/11 and grant No.~2022/WK/12.

\end{document}